\documentclass[a4paper,fleqn]{cas-sc}

\usepackage[authoryear]{natbib}
\usepackage{graphicx} 
\usepackage{float}
\usepackage{algorithm}  
\usepackage{algpseudocode}
\usepackage{color}
\usepackage{setspace}
\usepackage{enumitem}
\usepackage[nomarkers,figuresonly]{endfloat}

\def\tsc#1{\csdef{#1}{\textsc{\lowercase{#1}}\xspace}}
\tsc{WGM}
\tsc{QE}
\tsc{EP}
\tsc{PMS}
\tsc{BEC}
\tsc{DE}
\begin{document}
\let\WriteBookmarks\relax
\def\floatpagepagefraction{1}
\def\textpagefraction{.001}
\shorttitle{SAIPy and a Multi-Station-Based Approach}
\shortauthors{Quinteros-Cartaya et al.}

\title [mode = title]{Evaluating the SAIPy Performance using a Local Seismic Network for Volcano-Tectonic Earthquakes Monitoring}

%type=editor,auid=000,bioid=1,
\author[1]{Claudia Quinteros-Cartaya}[orcid=0000-0003-3487-2889]
\credit{Conceptualization, data processing and preparation, methodology, formal analysis and investigation, validation, original draft preparation, writing - review and editing}

\author[2]{Francisco Javier Núñez-Cornú}[orcid=0000-0003-1515-1349]
\credit{Data preparation, validation, writing - review and editing}

\author[1,3]{Nishtha Srivastava}[orcid= 0000-0003-0328-0311]
\credit{Conceptualization, funding acquisition, methodology, writing - review and editing}

\address[1]{Frankfurt Institute for Advanced Studies, Ruth-Moufang-Straße 1, Frankfurt am Main, 60438, Hessen, Germany.}
\address[2]{CA-276 UDG, Universidad de Guadalajara. Avenida Universidad 203, Puerto Vallarta, 48280, Jalisco, Mexico.}
\address[3]{Institute of Geosciences, Goethe University, Altenhöferallee 1, Frankfurt am Main, 60438, Hessen, Germany.}

\begin{abstract}
In this study, we evaluated the performance of SAIPy, an open-source Python package for deep learning-based seismic data analysis, by applying its single-station monitoring tools and extending its use to a seismic network based approach, using data from a local seismic network deployed in a Caldera. Although the integrated models into SAIPy for earthquake detection, magnitude estimation, seismic phase picking, and P-wave polarity classification, were originally trained on tectonic signals, we assess their performance in a more complex seismic environment that includes volcano-tectonic events, along with signal interference from distant earthquakes.
We also demonstrate the advantages of integrating outputs using multiple stations to improve event detection. SAIPy was able to identify a significantly larger number of local events than those included in previously published catalogs. SAIPy demonstrated reliable phase picking and P-wave polarity estimation, particularly for local volcano-tectonic events, with some limitations observed in the magnitude estimation for complex volcanic signals. These results support the utility of SAIPy for processing continuous seismic data and suggest that future retraining using data with physically standardized units, removing instrumental response, and including data from more diverse seismic sources, could improve its generalization for magnitude estimation to complex scenarios and different seismic networks and sensor types.
\end{abstract}
 
%\begin{coverletter}

%Dear Editors-in-Chief,
%\newline
 
%please find the enclosed manuscript "..." which we are submitting for exclusive consideration for publication in Computers \& Geosciences. We confirm that the submission follows all the requirements and includes all the items of the submission checklist.  
%\newline
 
%The manuscript presents ... 
%\newline

%We provide the source codes in a public repository with details listed in the section "Code availability".
%\newline

%Thanks for your consideration. 
%\newline

%Sincerely,
%\newline

%Authors names

%Corresponding author affiliation and e-mail
%\newline

%\textbf{Delete before submission:}

%Please confirm that your submission follows all the requirements of the guidelines, including the submission checklist:

%- Cover letter

%- Highlights

%- Authorship statement

%- The manuscript must be single column and double spaced

%- Reference must be in the author-date format

%- Code availability section 

%*The manuscripts that do meet the requirement guidelines will be desk-rejected.

%\end{coverletter}

%\begin{highlights}
%\item Highlight 1
%\item Highlight 2
%\item Highlight 3
%\item Highlight 4
%\item Highlight 5
%\end{highlights}

\maketitle 
\printcredits

\doublespacing
\section{Introduction}
\label{intro}

Accurate earthquake analysis is fundamental for both real-time seismic monitoring and processing previously recorded seismic data. These data contribute to the completeness and reliability of earthquake catalogs, which are essential for a wide range of seismological studies, including understanding seismic hazards, studying seismic source mechanisms, and monitoring earthquake activity over space and time.

Traditionally, seismic monitoring systems rely on signal processing algorithms such as short-time average to long-time average (STA/LTA) \citep{bib_sta-lta_det, bib_sta-lta_pick} for the tasks of event detection and phase picking. Although these methods are broadly used and well-established, their performance is often constrained by noise levels, parameter tuning, and their limited generalization to the diverse nature of seismic sources.

In recent years, deep learning techniques have emerged as powerful alternatives for seismic signal analysis, offering automated solutions to traditionally manual or parameter-sensitive tasks. By learning hierarchical features directly from waveform data, these models have shown superior performance, particularly in scenarios with low signal-to-noise ratios and large datasets \citep{bib_DLSeis}.

Among the tools developed from these advances is SAIPy, an open-source Python package that integrates deep learning techniques into a unified pipeline for seismic data analysis \citep{bib_saipy}. SAIPy was originally developed to operate on continuous seismogram data, particularly for small to moderate local events, through pretrained models for earthquake detection, magnitude estimation, seismic phase picking, and first-motion polarity classification. More recently, a second branch was included by \cite{bib_mages} and \cite{bib_deteq}, for the detection and magnitude estimation of large earthquakes using HR-GNSS data.

The branch corresponding to the seismogram analysis pipeline, which is the focus of this study, comprises three core deep learning models: CREIME\_RT, which performs event detection and magnitude estimation; DynaPicker\_v2 , which identifies P- and S-phase arrival times; and PolarCAP, which classifies the first-motion polarity of detected events \citep{bib_saipy, bib_polarcap, bib_dynapicker, bib_creime}. These models have been trained on open access benchmark datasets such as STEAD \citep{bib_stead} and INSTANCE \citep{bib_instance} and are adaptable to user-specific data through retraining.

In its original design, SAIPy focuses on processing continuous waveform data from individual seismic stations. Although this approach enables efficient analysis, it does not directly account for the spatio-temporal coherence of seismic signals recorded across a network of stations.

Multi-station analysis, especially in regions with dense seismic networks, can significantly enhance event association, reduce false positives, and improve estimates of event location and magnitude.

In this study, we develop and test a framework that leverages SAIPy’s modular outputs, such as arrival times and magnitudes, to evaluate its performance using clustered outputs from multiple stations to identify common events and support a more comprehensive analysis.

\section{Seismic Network-Based Approach}

Since SAIPy processes data independently per station, the integration of information across a network allows to associate detections that correspond to the same seismic event. However, associating detections is more complex than a clustering approach based solely on the temporal difference of P-wave arrival times.

Seismic detections clustering often include closely spaced arrivals that may be temporally aligned, but do not necessarily correspond to the same seismic event. This ambiguity can arise due to several factors, including seismic swarms, where numerous small events occur in close spatial and temporal proximity, as well as local triggering effects or high-amplitude impulsive noise generated by instrumental, anthropogenic, or environmental sources that can resemble true seismic phases and result in misleading detections.

A range of studies has addressed the challenges of earthquake detection by seismic networks using both traditional and machine learning-based approaches. Recent methods such as REAL \citep{zhang2019real} enable rapid phase association and event location based on P and S picks and travel-time consistency, while \citep{bergen2018fast} extended single-station waveform similarity measures to network-based detections by pairwise pseudo-association. Furthermore, several deep learning frameworks have been developed to jointly perform phase picking and event association such as the end-to-end architectures proposed by \cite{zhu2022endtoend} and the graph neural network approach introduced by \cite{xu2024gnn} which also include simultaneously earthquake location.

In this work, we propose a time-based P arrival association method that incorporates temporal coherence in arrival patterns throughout the station network. We first define an initial pre-clustering step that groups detections from unique stations if they satisfy a temporal proximity threshold $\Delta t$, relative to the earliest detection within the group.

The threshold $\Delta t$ is chosen based on the expected maximum travel-time differences across the network array geometry. This is particularly relevant because SAIPy is designed to work with local events (epicentral distances < 100 km), where only local networks are able to well record events and arrivals across the network are expected to occur relatively close in time. In practice, $\Delta t{\scriptstyle{max}}$ is set by the user based on the geometry of the network and the assumed minimum P-wave velocity (e.g., 5–6 km/s).

Formally, for each detection $d_k = (t_k, s_k)$, where $t_k$ is the detection time and $s_k$ is the station, we iterate over existing clusters $C_m$ and add $d_k$ if:
\begin{equation}
    |t_k - t^{(0)}_m| \leq \Delta t \quad \text{and} \quad s_k \notin \{ s_j \mid d_j \in C_m \},
\end{equation}
where $t^{(0)}_m$ denotes the time of the first detection in the cluster $C_m$. If no existing cluster satisfies these conditions, a new cluster $C_{\text{new}} = \{ d_k \}$ is created.

Following pre-clustering, we address overlapping detections shared between clusters and ambiguous associations, which are common during periods of high detection density, such as during swarms. We leverage the assumption that, during a swarm, events are spatially concentrated and therefore produce a consistent relative order of station arrivals. This assumption implies that the detection pattern, that is, the sequence of stations ordered by arrival times, remains approximately preserved among events from the same source region.

Given a set of overlapping clusters $\{C_1, C_2, \dots, C_n\}$, we compute the station pattern for each cluster:
\begin{equation}
    \text{Pattern}(C_i) = (s_1^{(i)}, s_2^{(i)}, \dots, s_{k_i}^{(i)}),
\end{equation}
where $t_{s_1^{(i)}} < t_{s_2^{(i)}} < \cdots < t_{s_{k_i}^{(i)}}$.

We then compare these patterns with the most frequently observed pattern in all clusters, denoted $P^*$, and assign a score to each cluster based on the number of matching stations:
\begin{equation}
    \text{Score}(C_i) = \left| \text{Pattern}(C_i) \cap P^* \right|.
\end{equation}

The cluster with the highest pattern match score is selected as the representative cluster for that connected component. This pattern-based clustering refinement is applied only when the detections are shared among others.

\section{Experiment}

\subsection{Dataset}
\begin{figure}
\centering
\includegraphics[width=0.75\textwidth]{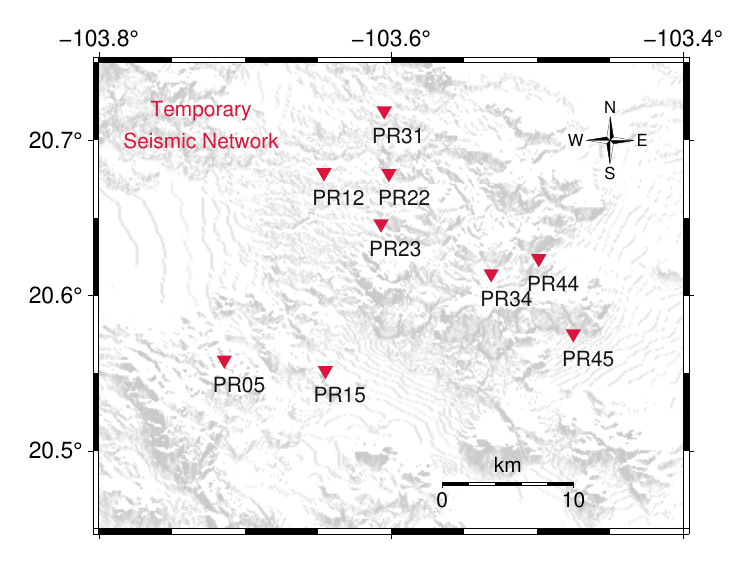}
\caption{ Seismic stations used in this work, which were part of a temporary network operative during 2017, in the framework of the P24 CeMIE-Geo project \citep{bib_PRIM_AGU, bib_GMZ}.}
\label{fig:Figure1}
\end{figure}

For this experiment, we used data from nine seismic stations that were part of a temporary local network (Figure \ref{fig:Figure1}). Since this network was deployed within a caldera that remains geothermally and seismically active, and also lies near an active rifting, it was able to record a variety of seismic events, including tectonics, volcano-tectonics, and volcanics. The stations were distributed locally so that the maximum distance between the stations does not exceed 30 km, ensuring adequate coverage for the detection of local and lower-magnitude events.

In this study, we focus on a short time window that we consider useful for testing and evaluation purposes. Specifically, we analyzed a few hours of data recorded on September 8 and 9, 2017; two days during which the seismicity rate was notably higher compared to other days in the dataset. As a reference, we based our analysis on the earthquake catalog compiled by \cite{bib_PRIM_AGU}.

\subsection{Implementation Details}
%as indicated in the package structure (see Figure X)
This work uses the updated version of the SAIPy modules. The core deep learning models remain unchanged; no modifications or retraining have been applied to the architectures of CREIME\_RT, DynaPicker\_v2, or PolarCAP. However, several supporting modules were edited to improve workflow organization, data handling, and visualization of results. These adjustments were made within the internal code. Details on these modifications can be found in the supplementary material.

\subsubsection{Single Station Monitoring}
The first stage involves independent monitoring of individual seismic stations. In this step, the input data consist of seismic waveforms read into an ObsPy Stream object, from data that have been preloaded into Python \citep{bib_obspy}. This data loading step is intentionally flexible, so the user can customize it depending on the specific structure and format of their dataset.

Furthermore, to ensure compatibility with SAIPy’s deep learning models, input seismic streams must be prepared in the same way as the data used during model training. That is, the models were trained on seismograms in three components (in strict order E, N, Z), in counts,  sampled at 100 Hz, and bandpass filtered between 1 and 45 Hz. Therefore, the input data must be ordered, resampled, and filtered accordingly. This preprocessing can be performed using the predefined module available within SAIPy, or alternatively, users may implement their own routines.

In the single-station monitoring stage, it is necessary to define a maximum time window length for earthquake analysis, which is used by the DynaPicker\_v2 model during the seismic phase picking process. This window determines the length of the signal segment analyzed from the detected time event. In this study, we set the window length to 15 seconds, considering the typical duration of the waveforms of the local events. This parameter can be defined empirically or through preliminary experiments, depending on the type of seismicity being studied.

\begin{figure}
\centering
\includegraphics[width=1\textwidth]{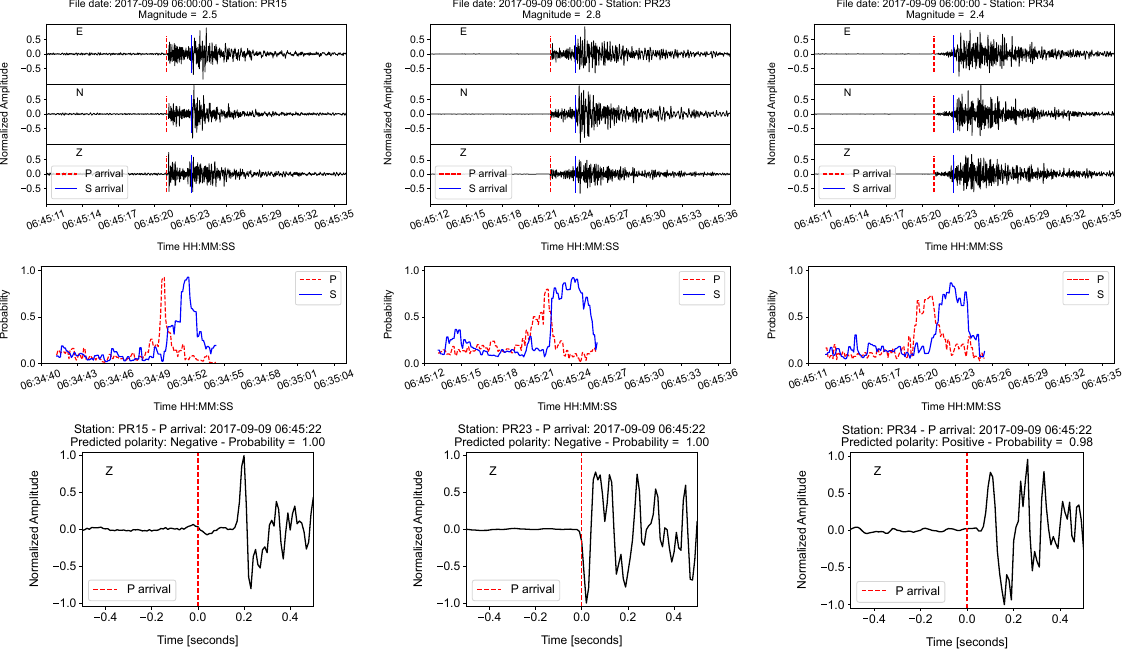}
\caption{ Output plots from the single-station monitoring by SAIPy. Example of three stations PR15, PR23, and PR45. This earthquake reported by \cite{bib_PRIM_AGU} as Ml 2.5, occurred on September 9, 2017, at 06:45:20, and 2 seconds later was recorded by these stations. It is worth noting that the reported location of this earthquake is located approximately midway between these three stations, just in the Caldera area. In the row of the top, are the P and S picking, and the magnitude estimated per station. The average of the magnitudes is very close to the reported Ml. In the middle row are the probabilities of the P and S picking, and in the bottom row, we can observe the P-arrival time and the respective polarities.}
\label{fig:Figure2}
\end{figure}

The results of this monitoring are stored in folders organized by station, with subfolders corresponding to specific dates and times. After this part of the analysis, it is possible to visualize the outputs, including plots of phase picking, phase pick probabilities, and P-wave polarity (Figure \ref{fig:Figure2}). Additionally, a CSV format file is generated for each station, summarizing all the parameters obtained for every detected event.

To proceed with the analysis of multiple stations and the event detection by P arrival association, the results are stored in a dictionary-formatted output, which facilitates structured access during the multi-station-based analysis.

\subsubsection{Detection by Multiple Stations}

Clustering of detections by the P arrival association requires the definition of a minimum number of stations that must report a detection to consider it as part of a valid cluster, as well as a maximum time difference $\Delta t{\scriptstyle{max}}$ between P-wave arrival times between stations to ensure that clustered detections likely correspond to the same seismic event. In this study, we used a minimum of two stations and a $\Delta t{\scriptstyle{max}}$ of 5 seconds. In the case of only two stations, the results are useful only for testing purposes; it is important to note that three or more stations would subsequently be necessary to obtain a reliable location of the event.

\begin{figure}
\centering
\includegraphics[width=0.6\textwidth]{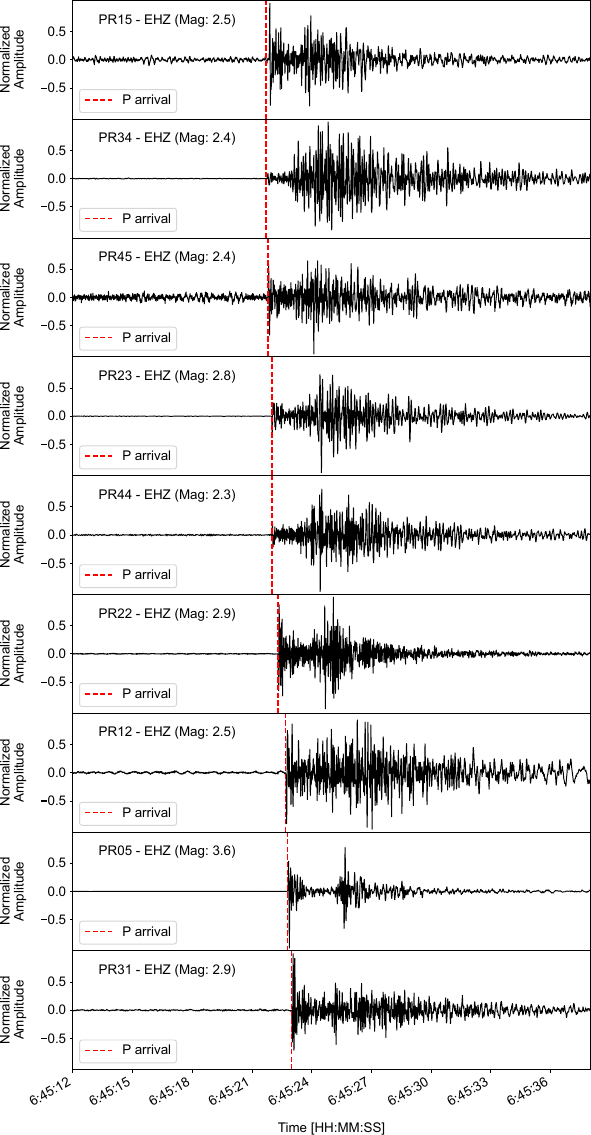}
\caption{ Output plot from the detection by multiple stations.  Only vertical components of the seismogram are shown. Example using the same earthquake as in Figure \ref{fig:Figure2}, occurred on 9 September 2017, which was detected by the nine stations used in the testing, sorted by detection time from the top to the bottom.}
\label{fig:Figure3}
\end{figure}

After completing the analysis, a CSV format file is generated in a designated output folder, containing information about the events identified, organized by date and time (depending on the time window of the input data, for example, hourly or daily). For each detected event, the Z-component waveforms per station can be visualized (Figure \ref{fig:Figure3}). In addition, all information related to the detections is stored in dictionary-formatted outputs for posterior creation of customized plots and further analysis (Figures \ref{fig:Figure4}).

\subsection{Performance Evaluation}

There are several important aspects to consider when analyzing the results of these tests. First, the dataset includes signals not only from tectonic sources but also from volcanic activity. In fact, a volcanic swarm can be observed during this period. This is an important consideration, as the models used in SAIPy were originally trained on local tectonic signals only, making this a good scenario to test their generalization capabilities. Furthermore, during the time window of the data analyzed in this work, aftershocks related to the Mw 8.2 Tehuantepec earthquake, which occurred in the early hours of 8 September, were still taking place (Figures \ref{fig:Figure5a} and \ref{fig:Figure5b}). Although the mainshock was around 1000 km away from these stations and the low-frequency signals recoded from this large earthquake and its aftershocks were mostly removed by the applied bandpass filtering, residual signals from these distant events could still be present, interfering with the local seismic signals.

Nevertheless, the integration of deep learning models with a multi-station network approach, as explored in this study, demonstrates strong potential to improve seismic event detectability and, consequently, enhance the completeness of seismic catalogs. For example, on 8 September between 05:00 and 06:00 (Figure \ref{fig:Figure5a}), six events were identified using the multi-station SAIPy analysis (Table \ref{tab1}), whereas only two of these were previously reported by \cite{bib_PRIM_AGU}. Similarly, between 17:00 and 18:00 (Figure \ref{fig:Figure5b}), 32 seismic events were detected, 19 of which were recorded by at least three stations (Table \ref{tab2}), and of these, only four were reported in the previous catalog by \cite{bib_PRIM_AGU}.

\begin{figure}
\centering
\includegraphics[width=1\textwidth]{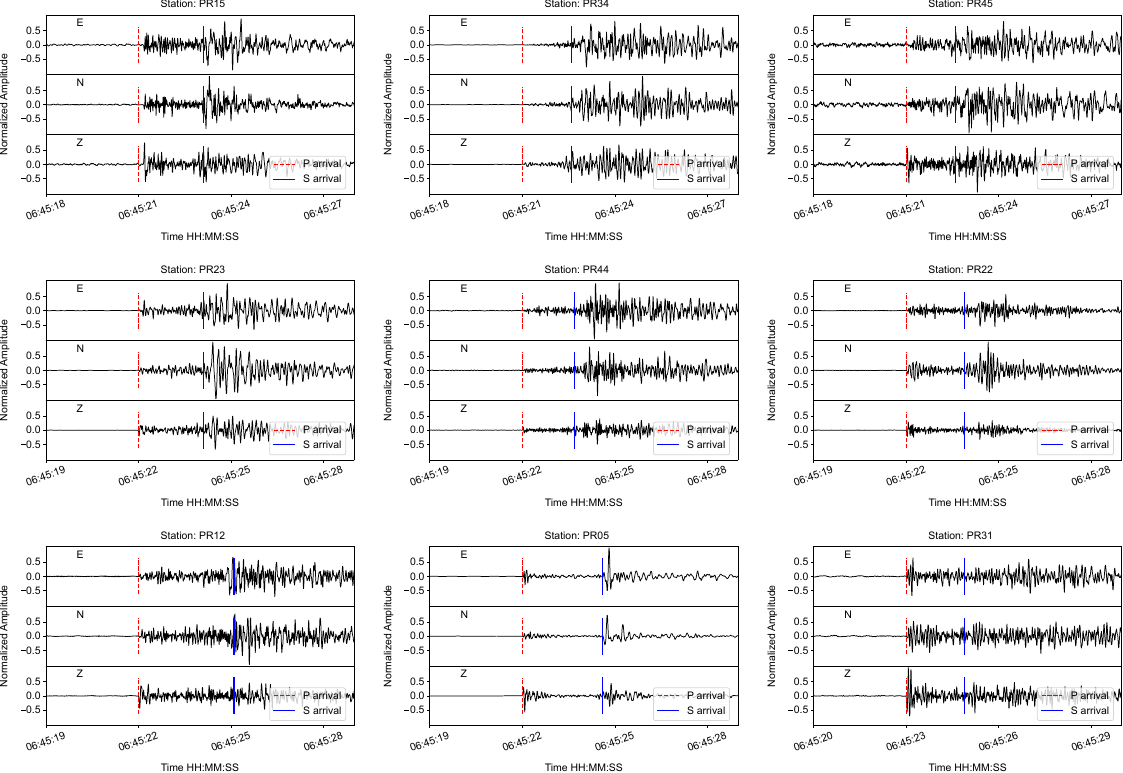}
\caption{Close-up view of the three components of the seismograms, with the P and S phase picks by SAIPy, of the event detected in nine stations showed in \ref{fig:Figure3}. In particular, this is an volcano-tectonic event located in the Caldera are. We could observe a very well Phase picking with differences of P-S arrivals no longer than 3 seconds. This visualization was customized to facilitate the review of picking accuracy.}
\label{fig:Figure4}
\end{figure}

\begin{figure}
\centering
\includegraphics[width=1\textwidth]{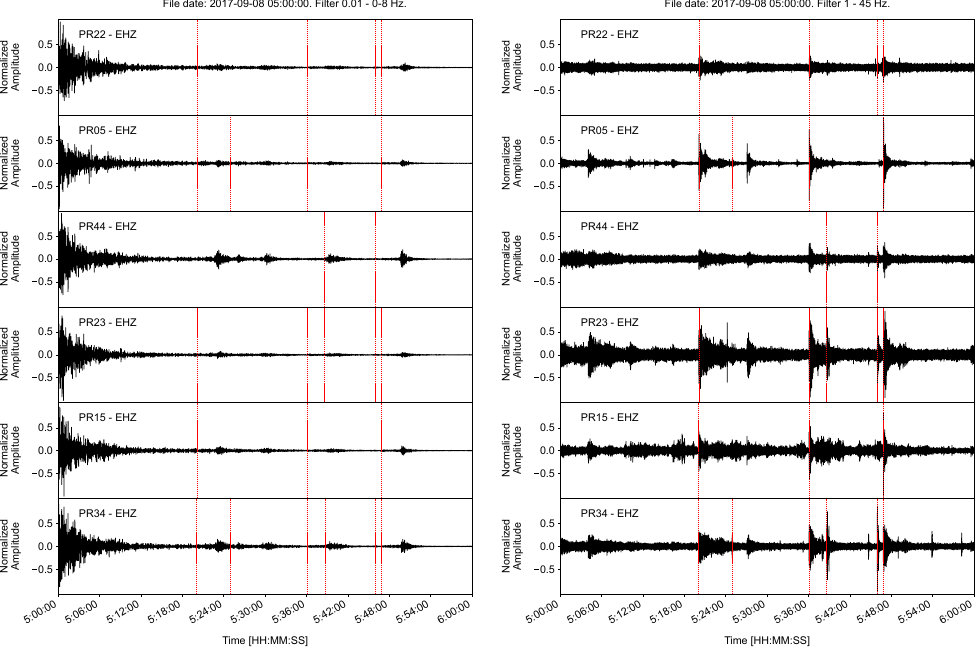}
\caption{One hour waveform recordings at six stations of the temporary network on September 8, 2017, starting at 05:00:00. Only vertical components of the seismogram are shown. In the left, the data are filtered to highlight low frequencies between 0.01 and 0.8 Hz, allowing the observation of regional earthquake signals, including aftershocks of the Mw 8.2 Tehuantepec earthquake that occurred at 04:49:17 (UTC). In the right, the data are filtered in the 1–45 Hz frequency range, which corresponds to the operating range of SAIPy, revealing a sequence of closely spaced local earthquakes. For comparison purposes, red lines that indicate the events detected using SAIPy's multi-station approach (with detections at two or more stations) are shown in both panels. Only two of the local events, during this time period, were reported by \cite{bib_PRIM_AGU} (see Table \ref{tab1}).}. 
\label{fig:Figure5a}
\end{figure}

\begin{figure}
\centering
\includegraphics[width=1\textwidth]{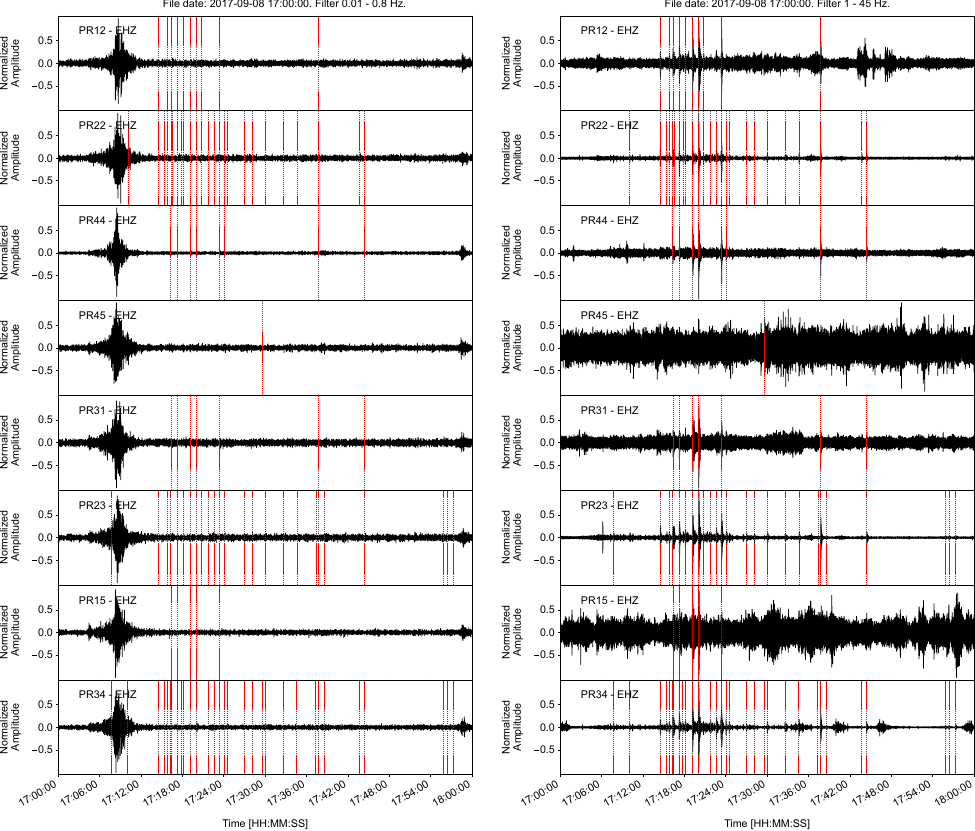}
\caption{One hour waveform recordings at eight stations of the temporary network on September 8, 2017, starting at 17:00:00. Only vertical components of the seismogram are shown. In the left, the data are filtered to highlight low frequencies between 0.01 and 0.8 Hz, allowing the observation of regional earthquake signals, including aftershocks of the Mw 8.2 Tehuantepec earthquake that occurred at 04:49:17 (UTC). In the right, the data are filtered in the 1–45 Hz frequency range, which corresponds to the operating range of SAIPy, revealing a sequence of closely spaced local earthquakes. For comparison purposes, red lines that indicate the events detected using SAIPy's multi-station approach (with detections at two or more stations) are shown in both panels. Only four of the local events, during this time period, were reported by \cite{bib_PRIM_AGU} (see Table \ref{tab2}).}. 
\label{fig:Figure5b}
\end{figure}

\begin{figure}
\centering
\includegraphics[width=1\textwidth]{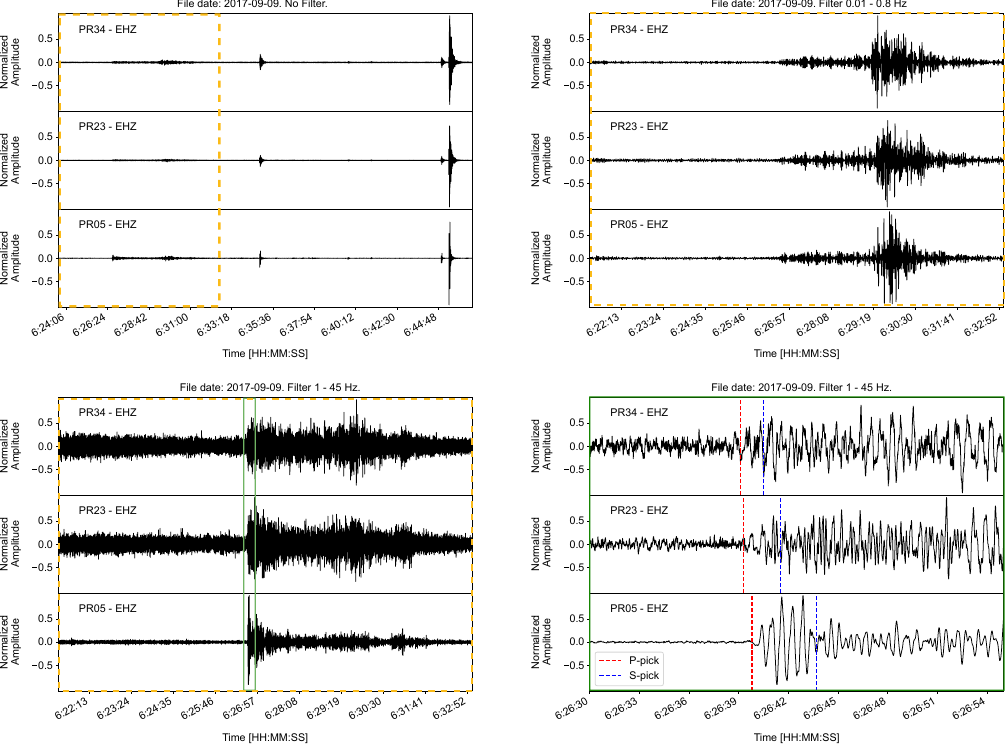}\caption{Example of the detection of a non-local earthquake using SAIPy. Only vertical components of the seismogram are shown. At the top left, 22 minutes of unfiltered recordings from three stations are shown. At the end of the trace, a high-amplitude event corresponding to the local earthquake previously shown in Figures 2, 3, and 4 can be observed. At the beginning of the trace, a regional earthquake is highlighted with yellow dashed lines. At the top right, the regional earthquake is shown filtered in the low-frequency band (0.01–0.8 Hz). At the bottom left, the same event is shown in the 1–45 Hz frequency range, which is used by SAIPy. At the bottom right, a close-up of the initial segment of the signal (outlined in green in the lower left panel) is displayed, showing the P- and S-wave picks as identified by SAIPy. Note that, since SAIPy is designed for local events and a fixed 15-second analysis window was used in this experiment, the model assigns an S-wave pick even when none is actually present within the analyzed time window.}
\label{fig:Figure6}
\end{figure}

\begin{figure}
\centering
\includegraphics[width=1\textwidth]{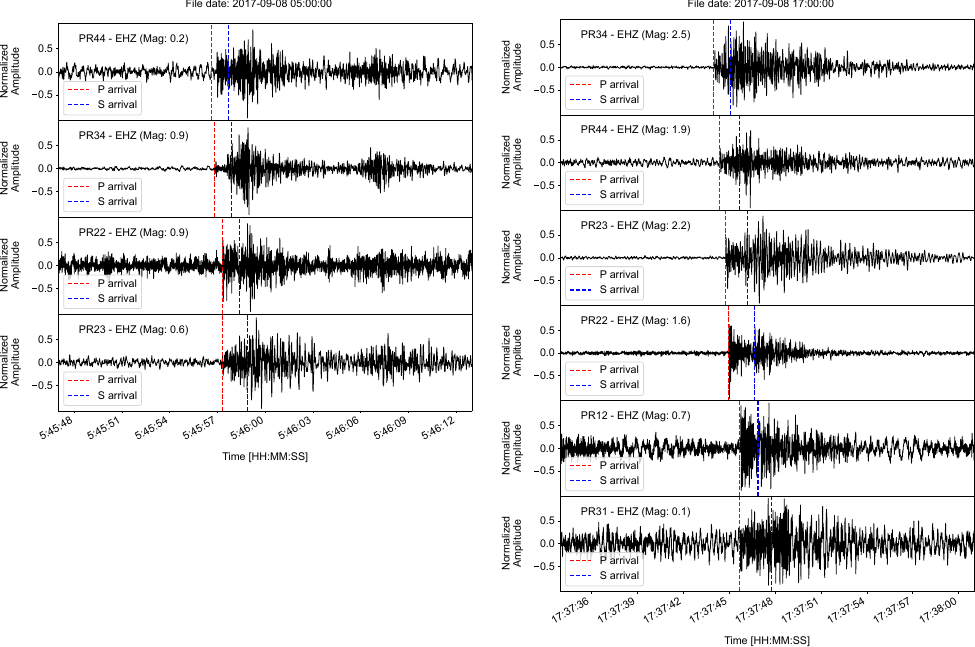}\caption{Examples of two events that occurred during the swarm on September 8, 2017. Only vertical components of the seismogram are shown. On the left, an event detected by four stations. On the right, an event detected at six stations. Both events were reported by \cite{bib_PRIM_AGU} as local volcano-tectonic earthquakes with Ml 1.5. The P- and S-wave arrival picks identified by SAIPy.}
\label{fig:Figure7}
\end{figure}

Several further observations are outlined below:
\begin{enumerate}[label=\alph*),ref=\alph*]

    \item \textbf{Detection by P arrivals association:} In our analysis, for this particular network where stations are spaced relatively close (within ~30 km maximum), the maximum expected P-wave travel-time difference between stations of $\Delta t_{\text{max}} = 5$ seconds provides a sufficient buffer to cluster detections from the same event. The choice of $\Delta t_{\text{max}}$ is sensitive to both network geometry and seismicity type. In case of swarms, supplementing time-based clustering with a pattern-matching step based on the sequence of station arrivals helps to distinguish overlapping detections from distinct events. However, in more spatially extended networks, when multiple events occur in rapid succession and no necessarily in close spatial proximity, fixing a proper $\Delta t_{\text{max}}$ becomes more challenging.
    
    \item \textbf{Time window length for phase picking:} We selected a 15-second window, which proved to be long enough to capture local earthquakes individually, even during seismic swarms (Figure \ref{fig:Figure5a} and \ref{fig:Figure5b}), where the inter-event time can be just a few minutes or seconds. At the same time, it is not too short, allowing both the P and S phases of local events to be observed within that window (e.g. Figure \ref{fig:Figure4}).
    
    However, some detections correspond to distant earthquakes, for which the waveforms are not fully contained within the 15-second window. As a result, the S-phase arrival is often incorrectly picked (Figure \ref{fig:Figure6}). A similar issue can arise with volcanic tremors, where the model, trained to always identify both P and S phases, attempts to pick phases even when the S phase is missing or not clearly present. This highlights the importance of reviewing visual outputs to manually exclude from the results such particular cases.
    
    \item \textbf{Magnitude Estimations:} In many cases, magnitude estimates at individual stations are consistent across the network; however, there are instances where certain stations produce outlier values, likely due to site effects or instrumental anomalies, for example, at station PR05 in Figure \ref{fig:Figure3}. In practice, these outliers are either excluded from average estimations or have reduced influence when the final result is based on a weighted average.

    Despite the fact that the CREIME\_RT model was trained and tested using data from a variety of instruments, some bias may still persist, particularly when applied to stations with instrumental responses or complex local site conditions that were not well represented in the training dataset. This issue is evident in cases showing notable dispersion in magnitude estimates among stations (Figure \ref{fig:Figure7}). In Tables \ref{tab1} and \ref{tab2}, we present the mean magnitude values and their respective standard deviations for events analyzed by SAIPy, compared with the local magnitudes reported by \cite{bib_PRIM_AGU}. Differences of up to 0.9 magnitude units are observed (e.g., Event ID 5 in Table \ref{tab1}), and in some cases, the mean magnitude matches the reported value, but with a standard deviation as high as 0.9 (e.g., Event ID 26 in Table \ref{tab2}). The corresponding phase picks and station-by-station magnitude estimates for both cases are shown in Figure \ref{fig:Figure7}.
    
    One important contributing factor could be that the CREIME\_RT model was trained using the local magnitude scale. The relationship between signal features and local magnitude is often biased by specific conditions such as attenuation and site effects, which may hinder the ability of the model to generalize.
    
    Particularly, in volcanic areas, characterized by heterogeneous subsurface structures, variable attenuation, and strong site effects, seismic wave propagation can vary substantially, especially given that most events are shallow and of very low magnitude \citep{bib_Havskov_Volc_Mag, bib_Zobin_Volc_Hazd}. These factors can lead to inconsistent signal features for the same event (e.g., amplitudes and frequencies) observed at different stations, further complicating the magnitude prediction process.
    
    Another limitation may stem from the fact that CREIME\_RT was trained using raw waveform counts rather than physical units (e.g., velocity or acceleration), introducing additional uncertainty due to amplitude inconsistencies related to the specific effects of the instrument.

\begin{table}
\centering
\caption{Magnitude of the events detected by SAIPy in a block of 1 hour, on 8th September 2017 from 05:00:00. In Italic are those previously reported by \cite{bib_PRIM_AGU}}.
\label{tab1}
\begin{tabular}{cccccc}
\toprule
\multicolumn{2}{c|}{} & \multicolumn{1}{c|}{\textit{Reported}} & \multicolumn{3}{c}{\textbf{SAIPy}} \\
\textbf{Event ID} & \textbf{Earliest P arrival} & \textit{Magnitude} & \textbf{Mean Magnitude} & \textbf{Std. Dev.} & \textbf{Nº Stations} \\
\midrule
1 & 05:20:04.1 & -   & 2.7 & 1.0 & 5 \\
2 & 05:24:52.8 & -   & 1.8 & 0.4 & 2 \\
3 & 05:36:03.3 & -   & 2.4 & 1.4 & 5 \\
4 & 05:38:37.3 & \textit{1.6} & 0.9 & 0.2 & 3 \\
5 & 05:45:56.6 & \textit{1.5} & 0.6 & 0.3 & 4 \\
6 & 05:46:50.1 & -   & 2.2 & 1.4 & 5 \\
\bottomrule
\end{tabular}
\end{table}

\begin{table}
\centering
\caption{Magnitude of the events detected by SAIPy in a block of 1 hour, on 8th September 2017 from 17:00:00. In Italic are those previously reported by \cite{bib_PRIM_AGU}}.
\label{tab2}
\begin{tabular}{cccccc}
\toprule
\multicolumn{2}{c|}{} & \multicolumn{1}{c|}{\textit{Reported}} & \multicolumn{3}{c}{\textbf{SAIPy}} \\
\textbf{Event ID} & \textbf{Earliest P arrival} & \textit{Magnitude} & \textbf{Mean Magnitude} & \textbf{Std. Dev.} & \textbf{Nº Stations} \\
\midrule
1 & 17:07:42.8 & -            & 1.0 & 0.0 & 2 \\
2 & 17:10:03.3 & -            & 1.0 & 0.1 & 2 \\
3 & 17:14:27.9 & -            & 1.3 & 0.4 & 4 \\
4 & 17:15:22.7 & -            & 1.4 & 1.2 & 2 \\
5 & 17:15:45.2 & -            & 0.5 & 0.3 & 4 \\
6 & 17:16:17.3 & \textit{1.7} & 1.0 & 0.7 & 7 \\
7 & 17:16:26.2 & -            & 1.8 & 0.2 & 2 \\
8 & 17:17:13.3 & -            & 0.8 & 0.7 & 7 \\
9 & 17:17:45.1 & -            & 0.6 & 0.6 & 2 \\
10 & 17:18:05.5 & -           & 0.7 & 0.6 & 4 \\
11 & 17:19:09.6 & \textit{1.6} & 1.1 & 0.6 & 7 \\
12 & 17:20:00.9 & -           & 1.7 & 0.5 & 7 \\
13 & 17:20:45.6 & -           & 0.5 & 0.6 & 3 \\
14 & 17:21:47.1 & -           & 1.4 & 0.6 & 3 \\
15 & 17:22:36.5 & -           & 1.6 & 0.8 & 3 \\
16 & 17:23:16.4 & -           & 0.9 & 0.3 & 7 \\
17 & 17:24:00.2 & -           & 1.8 & 0.3 & 4 \\
18 & 17:24:27.9 & -           & 0.7 & 0.4 & 2 \\
19 & 17:26:58.8 & -           & 0.8 & 0.9 & 3 \\
20 & 17:28:07.0 & -           & 1.4 & 1.0 & 3 \\
21 & 17:29:32.7 & -           & 1.2 & 0.1 & 2 \\
22 & 17:29:59.7 & -           & 0.7 & 0.2 & 3 \\
23 & 17:32:33.1 & -           & 0.8 & 0.7 & 3 \\
24 & 17:34:33.8 & \textit{1.5} & 1.1 & 0.1 & 3 \\
25 & 17:37:19.7 & -           & 0.2 & 0.5 & 2 \\
26 & 17:37:44.0 & \textit{1.5} & 1.5 & 0.9 & 6 \\
27 & 17:38:37.2 & -           & 0.2 & 0.4 & 2 \\
28 & 17:43:37.0 & -           & 1.1 & 0.1 & 2 \\
29 & 17:44:22.8 & -           & 1.2 & 0.9 & 5 \\
30 & 17:55:50.0 & -           & 0.6 & 0.2 & 2 \\
31 & 17:56:20.5 & -           & 0.7 & 0.1 & 2 \\
32 & 17:57:16.2 & -           & 0.3 & 0.3 & 2 \\
\bottomrule
\end{tabular}
\end{table}

    \item \textbf{Picking performance for tectonic vs. volcanic events:} In most cases, the DynaPicker\_v2 model provides accurate picks for both P- and S-wave arrivals (e.g., Figures \ref{fig:Figure4} and \ref{fig:Figure7}) in volcano-tectonic events, despite being originally trained only on tectonic signals. However, while the model is capable of analyzing events from different source types, the accuracy of S-wave picking can be more challenging in hybrid, low-frequency, or other volcanic events, where physical conditions, such as strong heterogeneity and the presence of magmatic fluids, can significantly affect S-wave propagation.
    
    \item \textbf{P-wave arrival polarities}: The final parameter provided by SAIPy is the P-wave polarity. When the P-phase picks are reliable and not significantly affected by noise, the corresponding P-wave polarities are generally reliable as well. The multi-station approach allows for clustering of these polarities, which could be useful for subsequent source mechanism analysis, particularly for events occurring within regions that have good azimuthal coverage by the seismic network. Even for non-local events, the PolarCAP model can yield good results, provided that the preprocessing is appropriate for those types of signals. 
\end{enumerate}

\section{Conclusions}

This study presents a practical implementation and evaluation of the SAIPy package for automated earthquake analysis using multiple stations from a local seismic network. In general, this work illustrates both the strengths and limitations of SAIPy, particularly in complex scenarios, such as the volcano-tectonic active region analyzed.

An important outcome of this study is the demonstration of SAIPy functionality using multiple stations. The respective outputs of single-station monitoring are clustered by the P arrival association to identify seismic events. This method significantly improves the robustness of the analysis, reducing false detections. Furthermore, previously undetected events were identified, demonstrating the ability of SAIPy to enhance the completeness of seismic catalogs.

Through this assessment, we identified notable aspects of SAIPy performance: the phase picking model performed reliably for local volcano-tectonic events. The magnitude estimations exhibited variability between stations, probably due to the volcanic nature of the events, the local complexities of the region, or instrumental effects. Future improvements could include retraining the models using data with instrumental effects removed and incorporating volcanic signals to further improve generalization and reduce the possibility of bias.

\section{Acknowledgments}

This research is supported by the ``KI-Nachwuchswissenschaftlerinnen'' - grant SAI 01IS20059 by the Bundesministerium für Bildung und Forschung - BMBF. Calculations were performed at the Frankfurt Institute for Advanced Studies GPU cluster, funded by BMBF for the project Seismologie und Artifizielle Intelligenz (SAI). Seismic data analyzed in this work were acquired in the framework of the P24 CeMIE-Geo project, “Passive seismic and magnetotelluric exploration of the Ceboruco and La Primavera volcanic fields”.

\newpage

\textbf{Code availability section}
The source codes are available via link: https://github.com/srivastavaresearchgroup/SAIPy

\bibliographystyle{cas-model2-names}
\bibliography{bibliography} 

\end{document}

\maketitle
\textbf{SAIPy Package – Updates Overview}

Here we provide a summary of changes since the first version and a list of new features, fixes, and structural changes.

\section{Structure of the SAIPy Project}

Below is the folder structure of the SAIPy package for seismogram analysis:

\begin{verbatim}
saipy/
|- __init__.py
|- user_settings.py             # Newly added
|- data/
|  |- __init__.py
|  |- base.py  
|  \- realdata.py
|- models/
|  |- __init__.py
|  |- creime.py
|  |- dinapicker.py
|  \- polarcap.py
|- modules/
|  |- __init__.py
|  |- multi_stations.py         # Newly added
|  |- phaseclassification.py
|  \- pytorchtools.py
\- utils/
|   |- __init__.py
|   |- networktools.py          # Newly added
|   |- packagetools.py 
|   |- picktools.py
|   \- visualization.py
\- gnss/                        # Branch for GNSS data analysis
\end{verbatim}

\section{New Additions}

\subsection*{user\_settings.py}
Includes user-defined functions related to data format and file name structures, used in SAIPy's modules to load data, label plots, and manage result storage:
\begin{itemize}[leftmargin=1.5em]
  \item \texttt{set\_filename} — must be defined by the user.
  \item \texttt{date\_format\_object} — must be defined by the user.
  \item \texttt{read\_non\_seismic\_format} — optional; only needed when using custom non-seismic data formats.
\end{itemize}

\subsection*{modules/multi\_stations.py}
Includes functions for event detection using multiple stations and the results are saved in a .csv file format. The clustering for the P arrival time association is based on temporal proximity of detections and arrival time pattern in different stations.

\subsection*{utils/networktools.py}
Contains:
\begin{itemize}[leftmargin=1.5em]
  \item \texttt{make\_monitor\_multistations}: to automatically load, pre-process, and analyze data of every station in the defined seismic network. The results are stored into a dictionary for a posterior event detection by network.
  
  \item \texttt{network\_detect}: to make the process for event detection by clustering of P arrival associated between multiple stations (minimum stations is set to 3 by default).
\end{itemize}

\subsection*{data/realdata.py}
New function:
\begin{itemize}[leftmargin=1.5em]
  \item \texttt{load\_streams}: Loads ObsPy streams for three components per station.
\end{itemize}

\subsection*{utils/packagetools.py}
New function:
\begin{itemize}[leftmargin=1.5em]
  \item \texttt{monitor}: Replaces \texttt{monitor1} and \texttt{monitor2} (functions in the version 1). This function works with a stream Obspy as input data.

\end{itemize}

\subsection*{utils/picktools.py}
New function:
\begin{itemize}[leftmargin=1.5em]
  \item \texttt{phase\_picking2}: Uses input data in numpy array format. No filtering is applied internally, since the data are preprocessed previously.
\end{itemize}

\subsection*{utils/visualization.py}
New plotting functions:
\begin{itemize}[leftmargin=1.5em]
  \item \texttt{plot\_data}
  \item \texttt{plot\_dynapicker\_output}
  \item \texttt{plot\_polarcap\_output}
  \item \texttt{plot\_multi\_stations}
  \item \texttt{plot\_overview}
\end{itemize}

\section{Modifications}

\subsection*{i) Preprocessing Improvements}
In \texttt{data/realdata.py}, preprocessing now includes:
\begin{itemize}[leftmargin=1.5em]
  \item Trend and mean removal
  \item Customized bandpass filtering (default: 1–45 Hz)
  \item Resampling to 100 Hz when required
\end{itemize}

\subsection*{ii) Function Renaming}
\begin{itemize}[leftmargin=1.5em]
  \item \texttt{utils/packagetools.py}
    \begin{itemize}
      \item \texttt{data\_windows} $\rightarrow$ \texttt{windows\_for\_creime}
      \item \texttt{polarity\_estimation} $\rightarrow$ \texttt{polarity\_classification}
    \end{itemize}
  \item \texttt{models/dynapicker.py}
    \begin{itemize}
      \item \texttt{load\_model} $\rightarrow$ \texttt{load\_dynapicker}
    \end{itemize}
\end{itemize}

\subsection*{iii) Path Requirement for Models}
The model path is now required as input for:
\begin{itemize}[leftmargin=1.5em]
  \item \texttt{CREIME\_RT} (\texttt{models/creime.py})
  \item \texttt{PolarCAP} (\texttt{models/polarcap.py})
  \item \texttt{load\_dynapicker} (\texttt{models/dynapicker.py})
\end{itemize}